\documentclass[epj]{svjour}

\usepackage{graphicx}
\usepackage{dcolumn}
\usepackage{bm}
\usepackage{units} 

\providecommand{\be}{\begin{equation}}
\providecommand{\ee}{\end{equation}}
\providecommand{\sub}[1]{_{\rm #1}}
\renewcommand{\sup}[1]{^{\rm #1}}
\providecommand{\refkl}[1]{(\ref{#1})}
\providecommand{\erw}[1]{\mbox{$\left \langle #1 \right \rangle$}} 
\providecommand{\abl}[2]   {\frac{d #1}{{\rm d} #2}}  
\providecommand{\diff}[1]{ \, d #1} 
\providecommand{\bdma}{\begin{eqnarray*}}
\providecommand{\edma}{\end{eqnarray*}}
\providecommand{\bdm}{\begin{displaymath}}
\providecommand{\edm}{\end{displaymath}}
\providecommand{\ablpart}[2]{\frac{\partial #1}{\partial #2}}  
\providecommand{\bi}{\begin{itemize}}
\providecommand{\ei}{\end{itemize}}

\providecommand{\ctrl}{r}

\providecommand{\fig}[2]{
   \begin{center}
     \includegraphics[width=#1]{#2}
   \end{center}
}

\providecommand{\entryTablePar}[2]{
     \hspace{0mm}\parbox{0.2\textwidth}{\vspace*{1mm}#1\vspace*{1mm}} \hspace*{-1mm}
  & \hspace{0mm}\parbox{0.2\textwidth}{\vspace*{1mm}#2\vspace*{1mm}} \hspace*{-1mm}
  \\  \hline
}

\begin{document}
\title{Approximate Hamiltonian Statistics in Onedimensional Driven Dissipative Many-Particle Systems}
\author{Martin Treiber\inst{1} \and Dirk Helbing \inst{2}
}
\titlerunning{Hamiltonian Statistics in Onedimensional Driven Dissipative Systems}
\authorrunning{Helbing and Treiber}
 
\offprints{treiber@vwi.tu-dresden.de} 
\institute{\inst{1} Institute for Economics and Traffic, 
  Dresden University of Technology,
  Andreas-Schubert-Str. 23, 01062 Dresden, Germany\\
\inst{2} ETH Zurich, UNO D11, Universit\"atstr. 41, 8092 Zurich, Switzerland}
\date{\today}
\abstract{
This contribution presents a derivation of the steady-state distribution of velocities and
distances of driven particles on a onedimensional periodic ring. We will compare two
different situations: (i) symmetrical interaction forces fulfilling Newton's law 
of ``actio = reactio'' and (ii) asymmetric, forwardly directed interactions as, for example
in vehicular traffic. Surprisingly, the steady-state velocity and distance distributions 
for asymmetric interactions and driving terms agree with the equilibrium distributions of
classical many-particle systems with symmetrical interactions, if the system is large enough. 
This analytical result is confirmed by computer simulations and
establishes the possibility of approximating the steady state
statistics in driven many-particle systems by Hamiltonian systems. Our
finding is also  useful to understand the various departure time
distributions of queueing systems as a possible effect of interactions
among the elements in the respective queue [D. Helbing et al., Physica
A 363, 62 (2006)].
} 

\maketitle

\section{Introduction}

Classical many-particle systems such as ideal gases are characterized by 
the applicability of Newton's laws of mechanics, which particularly includes
the law of ``actio = reactio''. With these basic laws, many fundamental 
properties can be derived, like the conservation of momentum and
energy. Such many-particle systems are known as Hamiltonian systems.
Even statistical physics and thermodynamics are based on these relationships.

But what would happen if the law of ``actio = reactio'' 
would not hold and the particle interactions would not fulfil momentum 
and energy conservation? One example of such a system are driven
Brownian particles where many results for stationary distributions are available.
However, in this system, interactions are taken into account
only implicitely by a nonlinear friction function \cite{erdmannEPJB2000}, or by a symmetric
interaction potential \cite{Haenggi-EPL1999}. In other systems
such as vehicular traffic one would like to model the non-symmetric
interactions between the particles (vehicles) explicitely.
Would it still be possible to find (analytical) formulas for the
stationary distributions of velocities and distances? In fact, although  a
statistical physics formalism for
driven systems is needed, there are still not many results
available. The existing results mainly concern the study of traffic-jam related 
condensation phenomena by means of the master equation \cite{Mahnke99}
or the Fokker-Planck equation \cite{Kuehne02}.

These considerations have assumed certain arrival and departure
rates to or from any forming vehicle clusters, but they have not
explicitly represented 
the acceleration or deceleration dynamics of interacting vehicles in
space and time, 
which may lead to dynamic self-organization phenomena 
such as emergent stop-and-go 
waves \cite{Helb-Opus}.  
In the following, we will take this dynamics
of interacting particles into  
account. 
\par
In fact, we are seeking for a method to treat dissipative driven many-particle systems in
a similar way as Hamiltonian systems. The idea is that the dissipation
in the system would 
be balanced by the effect of the driving force,
at least in a closed (circular) system in the limit of large particle numbers.
This idea has been used to evaluate the vehicle interaction potential
\cite{FPE-preprint,Krbalek_Helb,helbing2006uia},
but questioned to be applicable to systems with asymmetric, forwardly
directed interactions. 
Moreover, the method has been restricted to a very limited number of
potentials $U(s)$, as the 
normalization factor of the distance distribution could not be analytically determined in
general. 
\par
For onedimensional classical Hamiltonian
gases, many results have been previously derived in the framework of Random Matrix
Theory \cite{RMT-book,krbalek2001htf}. According to this, if coupled to a thermal
bath, the velocity distribution 
of gas particles is Gaussian and the distance distribution $g(s)$ can be written as 
\begin{equation}
  g(s) \propto \mbox{e}^{-[U(s)/\theta+Bs]} \, ,
\end{equation}
where $U(s)$ is the interaction potential, $\theta$ is the velocity variance
(i.e. proportional to the temperature), and $B$ depends on the
particle density. It would 
be very desireable to have a similar result for driven many-particle
systems, as this would 
allow one to determine the interaction potential and interaction force
among driven particles in 
the presence of fluctuations. Our hope is that, in the stationary state, the
dissipative interactions and the driving term would somehow cancel out
on average, so that the  
behavior would approximately correspond to a Hamiltonian system, as
assumed in the
Refs.~\cite{Krbalek_Helb,helbing2006uia,krbalek2001htf,krbalek2007edt,Schuetz-PRE2000}.
\par
 In fact, 
in this paper we will show that this idea is correct 
for onedimensional systems in the limit of
large particle numbers.
Even forwardly directed dissipative driven many-particle systems
behave approximately like Hamiltonian systems if they are
 far away from a dynamic instability.
\par
Our paper is structured as follows: In the next section, we lay out the
theoretical basis and derive the form of the onedimensional
Hamiltonian as well as the conditions, under which it provides a
correct description. In Sec. \ref{sec:predictions} we formulate the
predictions in a form that can be tested by simulating representative 
many-particle systems. The actual simulations and their results are
presented in  Section \ref{sec:results}, after which we conclude with a
discussion.

\section{\label{sec:theo}Driven many-particle model with dissipative interactions}

In the onedimensional driven-many particle system we discuss, point-like
particles $i$ change their location $x_i(t)$ in time $t$ according to the equation of motion
\begin{equation}
\label{SDEx}
  \frac{dx_i}{dt} = v_i(t) \, ,
\end{equation}
and their 
temporal velocity change $dv_i/dt$ is assumed to be given by the following
stochastic acceleration equation:
\begin{equation}
 \frac{dv_i}{dt} = \frac{v_0 - v_i}{\tau} + f(s_i) - \gamma f(s_{i-1}) + \xi_i(t) \, .
\label{SDEv}
\end{equation}
Here, $v_0$ denotes the ``free'' or ``desired'' velocity
and $\xi_i(t)$ represents a white noise fluctuation term
satisfying
\begin{eqnarray}
 \langle \xi_i(t) \rangle &=& 0 \, , \nonumber \\
\label{D}
 \langle \xi_i(t) \xi_j(t') \rangle &=& D \delta_{ij} \delta (t-t') \, ,
\end{eqnarray}
where $D$ is a velocity-diffusion constant.
The particle mass $m_i$ has been set to 1, and $f(s_i)\le 0$ describes a repulsive
interaction force, which depends on the particle distance 
$s_i(t) = x_{i+1}(t) - x_{i}(t)$.
The term $\gamma f(s_{i-1})$ with $0 \le \gamma \le 1$ allows to study different cases:
$\gamma = 1$ corresponds to the classical case of symmetrical interactions 
in forward and backward direction, fulfilling the physical law of ``actio~= reactio''.
$\gamma = 0$ corresponds to the case of forwardly directed
interactions only, which is, for example,
applicable to vehicles.

\subsection{Fokker-Planck equation for velocities and distances}

In order to determine the statistical distributions of the
velocities and distances of $n$ particles $i$, 
it is helpful to rewrite the above stochastic differential equation 
(Langevin equation) in terms of an equivalent Fokker-Planck equation. With the definitions
\begin{eqnarray}
\vec{s} &=& (s_1,\dots,s_n)\, , \quad \vec{v} = (v_1,\dots,v_n) \, , \\ 
 W(s_i,s_{i-1}) &=& v_0 + \tau [f(s_i) - \gamma f(s_{i-1})] \, , \nonumber \\
  \mbox{and } P &=& P(s_1,\dots,s_n,v_1,\dots,v_n,t) = P(\vec{s},\vec{v},t) \, , 
\end{eqnarray}
this Fokker-Planck equation reads \cite{Ri89}
\begin{eqnarray}
 \frac{\partial P}{\partial t} 
 &=&  \sum_{i=1}^n \bigg\{ - \frac{\partial}{\partial s_i} [ \underbrace{(v_{i+1} - v_{i})}_{=ds_i/dt} P] \nonumber \\
& & - \frac{\partial}{\partial v_i} \bigg[\underbrace{\bigg( \frac{W(s_i,s_{i-1}) - v_i}{\tau} \bigg)}_{=dv_i/dt - \xi_i} P \bigg] 
 + \frac{D}{2} \frac{\partial^2 P}{\partial v_i{}^2} \bigg\} , \quad
\label{FPG}
\end{eqnarray}
where we assume periodic boundary conditions $v_{k+n}(t) = v_k(t)$ and $s_{k+n}(t) = s_k(t)$
for a onedimensional ring of length $L$ with $n$ particles on it.
 In the following, we will 
show that the {\em ansatz} 

\begin{equation}
\begin{array}{l}
P(\vec{s},\vec{v}) = P(s_1,\dots,s_n,v_1,\dots,v_n)  \\
\quad = {\cal N} \exp\Big[-\sum\limits_j  \left(\frac{U(s_j)}{\theta} + B s_j 
 +\frac{(v_j - V)^2}{2\theta}\right) \Big]
\end{array}
\label{distr}
\end{equation}
is a stationary solution of the above Fokker-Planck equation, 
if the parameters $V$ and $\theta$
are properly chosen, and if the so-called interaction potential $U$
is defined by
\begin{equation}
\label{Ueff}
 U(s_i) = \frac{1+\gamma}{2} \int\limits_0^{s_i} ds \; f(s).
\end{equation}
In Eq. \refkl{distr},
\begin{equation}
 {\cal N} = \left[ 
  \int d^n s \int d^n v\; P(\vec{s},\vec{v}) \right]^{-1}
\end{equation}
is the normalization constant, and 
the Lagrange parameter $B$ is required to meed the constraint $\sum_i s_i = L$ 
determining the actual particle density. Moreover,
\begin{equation}
 V(t) = \langle v_i \rangle = 
 \int d^n s \int d^n v\; v_i P(\vec{s},\vec{v},t) 
\end{equation}
is the average velocity and
\begin{equation}
 \theta(t) = \langle (v_i -V)^2 \rangle = 
 \int d^n s \int d^n v\; (v_i - V)^2 P(\vec{s},\vec{v},t) 
\end{equation}
the velocity variance. In the following, we will restrict our investigation to the stationary case
with $dV/dt = 0$ and $d\theta/dt = 0$, which presupposes that the deterministic part of
Eq.~(\ref{SDEv}) fulfils the linear stability condition
\begin{equation}
(1-\gamma)^2 \frac{df(L/n)}{ds}  < \frac{1+\gamma}{2\tau^2} 
\label{linstab}
\end{equation}
(see Ref. \cite{EPJB2} for the method to determine this formula). 
Otherwise, dynamic patterns such as stop-and-go waves
may emerge from the dissipative interactions of driven particles \cite{Helb-Opus}. 
Notice that in the Hamiltonian case, $\gamma=1$, the stability
condition is always satisfied. Furthermore, the
factorization assumption (\ref{distr})  requires that all variables
$s_i$ and $v_i$ are statistically
independent from each other. 
According to numerical simulations,  this is only the case  if
$(1-\gamma)^2 df/ds$ is {\it much} smaller than the right-hand side of 
(\ref{linstab}), i.e., the system is {\it far} from the instability point. 
\par
With the ansatz \refkl{distr}, the three terms of the
Fokker-Planck equation \refkl{FPG}
can be written as 
\begin{eqnarray}
& & - \sum_i \frac{\partial}{\partial s_i} [(v_{i+1} - v_{i})P] \nonumber \\
&=& \sum_i (v_{i+1} - v_{i} ) \left[ \frac{1}{\theta} 
\frac{d U(s_i)}{d s_i} + B \right] P \nonumber \\  
&=& \sum_i (v_{i+1} - v_{i}) \left[ \frac{(1+\gamma)f(s_i)}{2\theta} +
B \right] P  \,, 
\end{eqnarray}
\begin{eqnarray}
& & - \sum_i \frac{\partial}{\partial v_i} \left( \frac{W(s_i,s_{i-1}) - v_i}{\tau} P \right) \nonumber \\
&=& \sum_i \frac{P}{\tau} - \sum_i \frac{W(s_i,s_{i-1}) - v_i}{\tau} \left[ - \frac{(v_i - V)}{\theta}\right] P \, , \qquad
\end{eqnarray}
and
\begin{equation}
 \sum_i \frac{D}{2} \frac{\partial^2 P}{\partial v_i{}^2} 
= \frac{D}{2} \sum_i \left[ - \frac{1}{\theta} 
+ \left( - \frac{v_i-V}{\theta} \right)^2 \right] P \, .
\end{equation}
We will now  use the fact that
\begin{equation}
\label{shift}
 \sum_i g_{i\pm 1}P = \sum_i g_i P
\end{equation}
for any $i$-dependent variable $g_i$, i.e. indices can be shifted because of the 
assumed periodic boundary conditions. In this way we find
\begin{eqnarray}
 \frac{\partial P}{\partial t} &=& \frac{1}{\theta}
 \sum_i  \frac{1+\gamma}{2} (v_{i+1} - v_{i}) 
f(s_i) P + n \left(\frac{P}{\tau}- \frac{DP}{2\theta} \right) \nonumber \\
&+& \frac{1}{\theta} \sum_i 
\left[ \frac{v_0 - v_i}{\tau} + f(s_i) - \gamma f(s_{i-1})\right] (v_i - V)  P \nonumber \\
&+& \frac{D}{2 \theta^2} \sum_i  (v_i - V)^2 P \, .
\label{last}
\end{eqnarray}
Remarkably, this equation does not depend on the Lagrange parameter $B$ anymore,
which is needed to adjust to the particle density.
\par
Note that ansatz~(\ref{distr}) can only be a stationary solution with $\partial P/\partial t = 0$,  if
\begin{equation}
\label{FDT}
 \frac{1}{\theta} = \frac{2}{D\tau} \, .
\end{equation}
This relationship corresponds to the fluctuation-dissipation theorem. Applying it also to the last
term of Eq.~(\ref{last})  and using the decompositions 
$(v_{i+1}-v_i) = (v_{i+1} - V) - (v_i - V)$ and $(v_0 - v_i) = (v_0 - V) - (v_i - V)$, we find
\begin{eqnarray}
 \frac{\partial P}{\partial t} 
&=& \sum_i \frac{1- \gamma}{2\theta}  (v_{i+1}+v_i-2V) f(s_i) P \nonumber \\
&+& \frac{1}{\theta} \sum_i \frac{(v_0 - V)(v_i - V)}{\tau}  P \, ,
\label{compare}
\end{eqnarray}
and,  with the factorization assumption \refkl{distr} and shifting
indices again according to \refkl{shift}, we obtain
\be
\label{FPG-result}
\ablpart{P}{t}=\frac{1-\gamma}{\theta}\sum
\left[f(s_i) + \frac{v_0 - V}{\tau}\right] \big(v_i - V \big) \, P.
\ee
We will distinguish the following cases:
\begin{itemize}
\item[1.] In the case of a classical many-particle system with momentum conservation ($\gamma = 1$) 
and energy conservation, i.e. no dissipation ($\tau \rightarrow \infty$), 
we find $\partial P/\partial t = 0$,
i.e. ansatz (\ref{distr}) is an exact stationary solution of the Fokker-Planck equation (\ref{FPG}).
\item[2.] In the case of forwardly directed interactions as in 
vehicle traffic ($\gamma = 0$) and vanishing correlations, we have 
\begin{eqnarray}
& &\lim_{n\rightarrow \infty} \frac{1}{n} \sum_i (v_{i}-V) \left[ f(s_i)   + \frac{v_0 - V}{\tau}\right] \nonumber \\
&=& \left[ \lim_{n\rightarrow \infty} \frac{1}{n} \sum_i (v_{i}-V) \right] \nonumber \\
& & \quad \times \left\{ \lim_{n\rightarrow \infty} \frac{1}{n} \sum_i \left[ f(s_i)   + \frac{v_0 - V}{\tau}\right] \right\} \, . 
\label{cal}
\end{eqnarray}
The first factor vanishes because of $V = \lim_{n\rightarrow \infty} \frac{1}{n} \sum_i v_i$,
but the second factor disappears as well:
Dividing Eq.~(\ref{SDEv}) by $n$ and summing up over $i$ gives
\begin{eqnarray}
 \frac{1}{n} \sum_i \frac{dv_i}{dt} &=& \frac{1}{n} \sum_i \frac{v_0 - v_i}{\tau}
+ \frac{1}{n} \sum_i f(s_i) \nonumber \\
&+& \frac{1}{n} \sum_i \xi_i(t) \, . 
\end{eqnarray}
In the limit $n\rightarrow \infty$ of large enough particle numbers $n$, 
the left-hand side converges to $dV/dt$, while the last term on the right-hand side converges
to 0. In the assumed stationary case with $dV/dt = 0$ and using
$v_0 - v_i = (v_0 - V) - (v_i - V)$, this implies
\begin{eqnarray}
 0 &=& \lim_{n\rightarrow \infty}
 \frac{1}{n} \sum_i \left[ \frac{v_0 - v_i}{\tau} + f(s_i) \right] \nonumber \\
 &=& \lim_{n\rightarrow \infty} \frac{1}{n} \sum_i \left[ \frac{v_0 - V}{\tau} + f(s_i) \right]
\label{because}
\end{eqnarray}
because of $\frac{1}{n} \sum_i v_i = V = \frac{1}{n} \sum_i V$.
\end{itemize}

\noindent
We conclude that the
factorisation ansatz \refkl{distr} satisfies the Fokker-Planck
equation \refkl{FPG} if either the momentum is conserved ($\gamma=1$),
 or if the single-particle
gaps and velocities are independent from each other.
Note that, in order to arrive at this conclusion, 
 the special form \refkl{Ueff} for the interaction potential, 
 particularly the prefactor
 $(1+\gamma)/2$,  is required.

\subsection{Hamiltonian description}

An alternative approach is the Hamiltonian description.
For this purpose, let us investigate the Hamiltonian
\begin{equation}
 {\cal H} = {\cal T} + {\cal V}
 = \sum_i \frac{(v_i - V)^2}{2} + \sum_i U(s_i) \, .
\end{equation}
If $dV/dt = 0$, we can derive the following relations:
\begin{eqnarray}
\frac{d{\cal H}}{dt} &=& \frac{d{\cal T}}{dt} + \frac{d{\cal V}}{dt} \nonumber \\
&=& \sum_i (v_i - V) \frac{dv_i}{dt}  + \sum_i \frac{dU(s_i)}{d s_i} \nonumber \\ 
& & \quad \times \left(\frac{ds_i}{dx_{i}}\frac{dx_{i}}{dt} + \frac{ds_{i}}{dx_{i+1}} \frac{dx_{i+1}}{dt} \right) \nonumber \\
&=& \sum_i (v_i - V) \frac{dv_i}{dt}  + \sum_i \frac{1+\gamma}{2} f(s_i) (v_{i+1} - v_{i}) \nonumber \\
&=& \sum_i (v_i - V) \left(  \frac{v_0 - v_i}{\tau} + f(s_i) - \gamma f(s_{i-1}) + \xi_i(t) \right) \nonumber \\
& & \quad + \sum_i \frac{1+\gamma}{2} f(s_i) (v_{i+1} - v_{i}) \nonumber \\
&=& \sum_i \frac{1-\gamma}{2} (v_i+v_{i+1}-2V) f(s_i)  \nonumber \\
& & \quad + \sum_i \frac{(v_0 - V)(v_i - V)}{\tau} \nonumber \\
& & \quad - \sum_i \frac{(v_i - V)^2}{\tau} + \sum_i (v_i - V) \xi_i(t) \, .
\end{eqnarray}
Comparing this with (\ref{compare}) shows that 
\begin{equation}
 \frac{\partial P}{\partial t} = \frac{P}{\theta} \frac{d{\cal H}}{dt} +\frac{1}{\theta} \sum_i \left[
   \frac{(v_i - V)^2}{\tau} - (v_i - V) \xi_i(t) \right] P \, .
\end{equation}
Correspondingly, in the stationary state $\partial P/\partial t = 0$ we have
\begin{eqnarray}
 \frac{d{\cal H}}{dt} &=& \sum_i \left[ (v_i - V) \xi_i(t) - \frac{(v_i-V)^2}{\tau} \right] \nonumber \\
&=& \sum_i (v_i -V)\left( \xi_i(t) - \frac{v_i - V}{\tau} \right) \, . 
\label{H}
\end{eqnarray}
In order to investigate under which conditions the Hamiltonian is
conserved in the statistical average, we will calculate
$\erw{\abl{{\cal H}}{t}}$ for two different cases:
\begin{itemize}
\item[1.] In a conservative system with no fluctuations ($\xi_i(t) = 0 = D$) and no
dissipation ($\tau \rightarrow \infty$), we have $d{\cal H}/dt = 0$, independently of whether
the interactions are symmetric or forwardly directed.
\item[2.] For many-particle systems with fluctuation terms and/or dissipation, one
can show
\begin{eqnarray}
 \langle \xi_i (v_i - V) \rangle &=& \left\langle \frac{1}{2} \frac{d(v_i - V)^2}{dt} \right\rangle
- \frac{v_0-V}{\tau} \langle v_i - V \rangle \nonumber \\
 & & \quad + \frac{1}{\tau} \langle (v_i -V)^2 \rangle \nonumber \\
 & & \quad - \langle [f(s_i) - \gamma f(s_{i-1})] (v_i-V) \rangle \nonumber \\
 &=& \frac{1}{2} \frac{d\theta}{dt} - \frac{v_0-V}{\tau}(\langle v_i \rangle - V)  + \frac{\theta}{\tau} \nonumber \\
& & \quad - \langle f(s_i) - \gamma f(s_{i-1}) \rangle (\langle v_i \rangle -V ) . \qquad 
\label{show}
\end{eqnarray}
This can be found by multiplication of Eq.~(\ref{SDEv}) with $(v_i-V)$
and calculation of the  
ensemble average, using the factorization ansatz (\ref{distr}). 
The first term on the right-hand side vanishes under the assumption of
a stationary state. 
The second and the fourth term vanish because of $\langle v_i \rangle
= V$. Therefore, 
\begin{equation}
 \langle \xi_i (v_i - V) \rangle = \frac{\theta}{\tau},
\end{equation}
and, together with Eq.~(\ref{H}), we arrive at 
\begin{equation}
 \qquad \quad \left\langle \frac{d{\cal H}}{dt} \right\rangle 
= \frac{1}{\tau} \left(n \theta - \sum_i (v_i - V)^2\right)
 = 0 \, .
\end{equation}
That is, in the statistical average we have $d{\cal H}/dt = 0$. The same is expected 
for the average Hamiltonian per particle, $H_1={\cal H}/n$ of
systems with many particles. In fact, simulations show that $H_1$
fluctuates with amplitudes $\propto 1/\sqrt{n}$, while the  Hamiltonian
${\cal H}$ itself fluctates with amplitudes $\propto \sqrt{n}$, which is
consistent with  equilibrium hydrodynamic systems.
As a consequence, stationary driven dissipative systems
behave approximately Hamiltonian, even if the interactions are forwardly directed and
Newton's law ``actio = reactio'' is violated. This is, why the Hamiltonian statistics
\begin{equation}
 P(\vec{s},\vec{v}) = {\cal N} \mbox{e}^{-{\cal H}/\theta} 
\label{ap}
\end{equation}
(the canonical distribution) is an approximate stationary solution of our driven dissipative many-particle system.
Note that the contribution $\sum_i B s_i = BL$ in Eq.~(\ref{distr}) 
gives just a constant prefactor and can be absorbed into
the normalization factor.
\end{itemize}
In conclusion, the equilibrium solution (\ref{distr}) of conservative
many-particle systems is also a 
good approximation for the steady-state solutions
($\partial P/\partial t =0$) 
of driven many-particle systems of 
kind (\ref{SDEv}) with asymmetrical interactions, driving and dissipation effects, if the system is
large enough, i.e. $n\gg 1$ 
(for small systems, we expect that fluctuations become essential), 
and if the correlations between the gaps and
velocities are insignificant. 

\section{\label{sec:predictions}Application to stochastic traffic models}
%
In this section, we will formulate  the results of the previous
section in a way that can be tested by means of
simulating specific models.

\subsection{Single-particle distributions}

The factorisation \refkl{distr} can be written in the form
\begin{equation}
P(s_1,\dots,s_n,v_1,\dots,v_n) = \prod_{i=1}^n g(s_i) \prod_{j=1}^n h(v_j) \, ,
\label{factor}
\end{equation}
i.e., the statistics of the particles can be described by the
single-particle gap distribution function
\begin{equation}
\label{gs}
 g(s) = A \mbox{e}^{-[U(s)/\theta + B s]},
\end{equation}
and the single-particle velocity distribution
\begin{equation}
\label{hv}
h(v) = \frac{1}{\sqrt{2\pi\theta}}\mbox{e}^{-(v - V)^2/(2\theta)}.
\end{equation}
Here, $A$ is a normalization constant, $B$ a Lagrangian parameter
ensuring the density constraint,
$V$ 
the average velocity, and $\theta$ 
the velocity variance. With the exception of $\theta$, all quantities 
are dependent on the particle density $\rho$.

\subsection{Gap distribution}
The two constants $A$ and $B$ of the gap distribution \refkl{gs} are
determined using the normalisation condition
\be
\label{norm-cond}
\int \limits_0^{\infty} \diff{s} \, g(s) = 1\,,
\ee
and the constraint  that the average gap is equal to the
inverse of the global 
density,
\be
\int \limits_0^{\infty} \diff{s} \ s \, g(s) = \frac{1}{\rho} \, .
\ee
Defining the integrals
\bdma
I_0(B) &=& \int \limits_0^{\infty} \diff{s}
    \ \exp\left[-\left(\frac{U(s)}{\theta}+Bs\right)\right],\\
I_1(B) &=& \int \limits_0^{\infty} \diff{s}
    \ s \exp\left[-\left(\frac{U(s)}{\theta}+Bs\right)\right],\\
I_2(B) &=& \int \limits_0^{\infty} \diff{s}
    \ s^2 \exp\left[-\left(\frac{U(s)}{\theta}+Bs\right)\right],
\edma
we get 
\be
A=\frac{1}{I_0(B)} \, .
\ee
For $B$, we find the transcendental  equation
\be
A I_1(B) =\frac{I_1(B)}{I_0(B)}=\frac{1}{\rho}.
\ee
Using Newton's method with the initial guess 
$B_0=\rho+1/\sigma_s$ with $\sigma_s$ defined in Eq. \refkl{sigs}, 
one obtains for the $k$-th iteration
\be
\label{Newton}
B_{k+1}=B_k + \frac{I_0(I_1\rho-I_0)}{\rho(I_2I_0-I_1^2)},
\ee
where the integrals on the right-hand side are evaluated at $B=B_k$. 
It turns out that this method converges  within very few iterations, unless 
$U(s_{\rm e})=U(1/\rho) \gg \theta$.
In this case, however, the second derivative $U''(s_{\rm e})$ 
of the effective potential \refkl{Ueff} typically satisfies the  condition
\be
\label{sigsGauss-condition}
\frac{|U''(s_{\rm e})|}{\theta} = \frac{(1+\gamma) |f'(s_{\rm e})|}{D\tau} \gg \rho^2\, , 
\ee
allowing an asympotic
expansion of \refkl{gs}  that eventually leads  to a
Gaussian
gap distribution,
\be
\label{sigsGauss}
\tilde{g}(s)_{|U''(s_e)|\gg \rho^2}
 =\frac{1}{\sqrt{2\pi}\sigma_s}e^{\frac{(s-s_e)^2}{2\sigma_s^2}}
\ee
with
\be
\label{sigs}
\sigma_s^2 = \frac{D \tau}{(1+\gamma)f'(s_e)}.
\ee
Remarkably, the ranges of applicability of \refkl{Newton} and
\refkl{sigsGauss} generally overlap, allowing a fast and robust solution.

\subsection{Velocity distribution and kinetic energy}

Equation \refkl{hv} states that, regardless of the density, of the potential, and of the
directions of the interactions,  the single-particle velocity distribution 
is Gaussian. The expectation value is equal to that of the stationary
velocity without fluctating terms. Furthermore, the velocity variance satisfies the 
fluctuation-dissipation theorem \refkl{FDT}, i.e., the average kinetic energy
per particle should be given by 
\be
\label{kin}
T=\frac{1}{2} \erw{(v_i-V)^2}=\frac{D\tau}{4}.
\ee

\section{\label{sec:results}Results}
In this section, we will show by means of computer simulations, 
 that the main predictions \refkl{gs}, \refkl{hv},
and \refkl{kin} are valid if the system is in a regime
that is far from any collective instability. 
In order to quantify this condition, we observe from 
\refkl{linstab} that  the system becomes linearly unstable if the
relaxation time $\tau$ exceeds some critical value $\tau_{\rm c}$. Thus,
$\tau$ controls the stability properties which allows us to define
the dimensionless reduced control parameter
\be
\label{ctrl}
\ctrl=\frac{\tau}{\tau_{\rm c}}.
\ee
Note that $\ctrl=0$ denotes equilibrium, while the linear threshold is
characterized by $\ctrl=1$. In particular, in the momentum-conserving
case $\gamma=1$, we have always $\ctrl=0$. The condition ``far
away from the instability point'' can be formulated by the condition
$\ctrl \ll 1$.

\subsection{\label{sec:models}Selected models}

In order to obtain a specific model, the interaction force of
the stochastic differential equation \refkl{SDEv} has to be specified. 
We will simulate two types of  interaction forces that are based
on (i) the optimal-velocity
model, and (ii) on a power law.

\begin{table}
\begin{center}
 \begin{tabular}{|l|l|}  \hline
\entryTablePar{Parameter}{Value} \hline
 \hline
\entryTablePar{Desired velocity $v_0$}{\unit[30]{m/s}}
\entryTablePar{Velocity relaxation time $\tau$}{\unit[0.2]{s}}
\entryTablePar{Interaction length $l\sub{int}$}{\unit[20]{m}}
\entryTablePar{Shape parameter $\beta$}{0.5}
\entryTablePar{Symmetry parameter $\gamma$}{0  and 1}
\entryTablePar{Fluctuating force $D$}{$\unit[20]{m^2/s^3}$}
\entryTablePar{Density $\rho$}{\unit[12]{/km} and \unit[30]{/km}}
\end{tabular}
\end{center}

\caption{\label{tab:paramOVM}Parameter values used in the simulations
for the stochastic OVM \protect\refkl{SDEv} with 
\protect\refkl{f-OVM} and \protect\refkl{ve_OVM}.
}
\end{table}

\begin{table}
\begin{center}
 \begin{tabular}{|l|l|}  \hline
\entryTablePar{Parameter}{Value} \hline
 \hline
\entryTablePar{Desired velocity $v_0$}{\unit[30]{m/s}}
\entryTablePar{Velocity relaxation time $\tau$}{\unit[2]{s}}
\entryTablePar{Interaction distance $l\sub{int}$}{\unit[20]{m}}
\entryTablePar{Acceleration $a_0$}{$\unit[2]{m/s^2}$}
\entryTablePar{Interaction exponent $\delta$}{2}
\entryTablePar{Symmetry parameter $\gamma$}{0 and 1}
\entryTablePar{Fluctuating force $D$}{$\unit[0.2]{m^2/s^3}$}
\entryTablePar{Density $\rho$}{\unit[10]{/km}}
\end{tabular}
\end{center}

\caption{\label{tab:paramPow}Parameter values used in the simulations
for the stochastic power-law model \protect\refkl{SDEv} with \protect\refkl{f-pow}.
}
\end{table}

\subsubsection{Stochastic optimal-velocity model (sOVM)}
In the \textit{stochastic optimal-velocity model} (sOVM), the interaction
force $f$ is given by \cite{helbing2006uia}
\be
\label{f-OVM}
f\sub{sOVM}(s)=\frac{V\sub{OVM}(s)-v_0}{\tau},
\ee
where we assume the optimal-velocity function
\be
\label{ve_OVM}
V\sub{OVM}(s)=\frac{v_0
\left[\tanh\left(\frac{s}{l\sub{int}}-\beta\right)+\tanh(\beta)\right]}
{1+\tanh(\beta)} \,.
\ee
Table \ref{tab:paramOVM} summarizes the meaning of the model
parameters and the values used in the simulations.
Notice that the conventional
optimal-velocity model of Bando et. al. \cite{Bando-jphys} is obtained
for the special case $\gamma=0$ and $D=0$ in the Eqs. \refkl{SDEv} and
\refkl{D}, respectively. 

The sOVM has the following properties:
The expectation value of the velocity for
stationary conditions is given by
\be
\label{V-OVM}
V\sub{sOVM}(s)= v_0+(1-\gamma)V\sub{OVM}(s).
\ee
Furthermore, the effective potential
\refkl{Ueff}  can be calculated analytically,  resulting in
\be
\label{U-OVM}
U\sub{sOVM}(s)=U_0 
\ln\left\{ 1+\exp\left[-2\left(\frac{s}{l\sub{int}}-\beta\right)\right]\right\} , 
\ee
where the prefactor is given by
\be
\label{U0}
U_0= \frac{(1+\gamma) v_0 l\sub{int}}
{2\tau \left(1+\tanh \beta \right)}.
\ee
The dynamics of this model becomes linearly unstable if the
relaxation time exceeds the critical value $\tau_{\rm c}$ given by
\be
\label{tauc-OVM}
\tau_{\rm c}(\rho)=\frac{1+\gamma}{2(1-\gamma)^2 V'\sub{OVM}(1/\rho)},
\ee
see Eq. (\ref{linstab}).
For the parameters specified in Table \ref{tab:paramOVM}, $\gamma=0$, and
$\rho=30$/km, the critical value is given by 
$\tau_{\rm c}(\gamma=0)=1/(2 V'\sub{OVM}(1/\rho)) =\unit[1.51]{s}$. This
corresponds to 
$\ctrl=0.132$ 
when the parameters of Table
\ref{tab:paramOVM} are assumed.

\subsubsection{\label{sec:power}Stochastic power law model (sPLM)}

An alternative, more physics-oriented model
assumes that the interaction
forces obey a power law  \cite{FPE-preprint,Krbalek_Helb}:
\be
\label{f-pow}
f\sub{sPLM}(s)=-a_0\left(\frac{l\sub{int}}{s}\right)^{\delta},
\ee
which, together with \refkl{SDEx} and \refkl{SDEv}, results
 in the stochastic power-law model (sPLM). The associated effective
potential \refkl{Ueff} is given by
\be
\label{U-pow}
U\sub{sPLM}(s)
 = \frac{(1+\gamma) a_0 l\sub{int}^{\delta}}
{(\delta-1)s^{\delta-1}},
\ee
and the 
expectation value for the velocity is equal to 
\be
\label{V-pow}
V\sub{sPLM}(s)=v_0 - (1-\gamma)\tau a_0\left(\frac{l\sub{int}}{s}\right)^{\delta} . 
\ee
This model differs qualitatively from the stochastic OVM in
the following aspects:
\bi
\item If $\gamma<1$, the stationary velocity becomes zero  for a finite
average gap,
$s_{\rm e}(V=0)=[(1-\gamma) \tau a_0/v_0]^{1/\delta}$. In contrast, 
 the stationary velocity of the sOVM for any
$\gamma>0$ is nonzero, even at maximum density, $s_{\rm e}=0$.
\item The potential \refkl{U-pow} of the stochastic power-law model diverges for
$s\to 0$, while the potential \refkl{U-OVM} of the sOVM remains
finite. For the chosen parameters and $\gamma=0$, we have 
$U\sub{sOVM}(0)$ = $U_0 \ln(2)$ = $\unit[1347]{m^2/s^2}$.
Consequently, any car
approaching a standing vehicle with a velocity exceeding 
$\sqrt{2 U\sub{sOVM}(0)}$ = \unit[52]{m/s} will lead to a rear-end
collision. This velocity decreases with increasing values of $\tau$,
reaching $\unit[18.9]{m/s}$ at the limit of linear stability. In contrast, no such
collisions are possible in the stochastic power-law model. Nevertheless, this
model can become
linearly unstable as well.
\ei

\subsection{Simulations}

We have simulated a closed ring road of length $L=\unit[9]{km}$ for
the sOVM, and $L=\unit[40]{km}$ for the sPLM.
We also have simulated larger systems resulting in no  significant
differences. We have started the simulations with deterministic
initial conditions $s_i=1/\rho$, and $v_i=V(s_i)$, corresponding to
a single-particle distribution function
\be
\label{IC}
P_1(s,v,0)=\delta \left(v-V\Big(\frac{1}{\rho}\Big) \right) 
\delta \left(s-\frac{1}{\rho}\right),
\ee
where $\delta(.)$ represents Dirac's delta function.
Since this initial condition does not correspond to a stationary
solution, we have run the  simulations for a transient time of \unit[72\,000]{s},
before  recording the results for further \unit[36\,000]{s}.

For the numerical update, we have applied the explicit scheme
\begin{eqnarray}
\label{update-v}
v_i(t+\Delta t) &=& v_i(t) + a_i(t)\Delta t + z_t \, \sqrt{D \Delta t}, \\
x_i(t+\Delta t) &=& x_i(t) + \left[\frac{v_i(t)+v_i(t+\Delta t)}{2}\right]\Delta t,
\end{eqnarray}
where $a_i(t)$ denotes the deterministic 
part of the right-hand side of Eq. \refkl{SDEv}, and $z_t \sim N(0,1)$ are
independent realisations of a Gaussian
distributed quantity with zero mean and unit variance.

The velocity update \refkl{update-v} corresponds to 
decomposing the deterministic and
stochastic parts of the accelerations.
While the deterministic part corresponds to an
Euler update, the stochastic part is a result of 
explicitely  solving the stochastic
differential equation
\bdm
\abl{v_i\sup{(s)}}{t'}= \xi_i(t')
\edm
for the initial conditions $v_i\sup{(s)}(t)=0$ at time $t'=t$. The
solutions $v_i\sup{(s)}(t+\Delta t)=\sqrt{D \Delta t}\ z_t$ are
realisations of random-walk trajectories, which are Gaussian distributed
 with expectation value zero,
and variance $D\Delta t$.

Notice that, for sufficiently small update
times, this update scheme should converge (in the statistical
sense)  to the true solutions of \refkl{SDEx} and \refkl{SDEv}.  
In the simulations, we have set  $\Delta t=\unit[0.04]{s}$. To verify
the convergence, we have also run
some simulations with lower values of the time step (down to $\Delta
t=\unit[0.001]{s}$)  and found less than 1\% deviation.

\begin{figure}
\fig{0.45\textwidth}{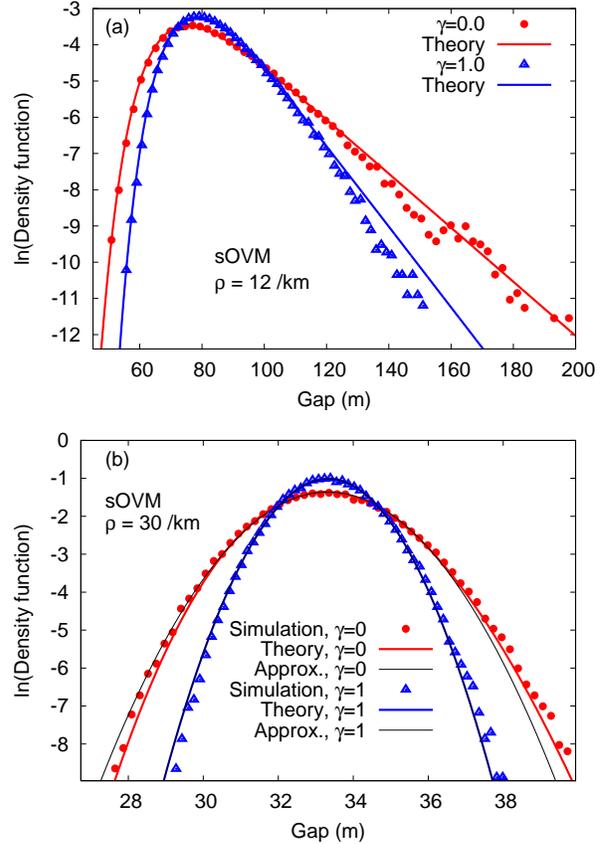}
\caption{\label{fig:OVM-s}Stationary gap distributions  for the
stochastic optimal velocity model (sOVM) with the parameters specified
in  Table \protect\ref{tab:paramOVM}
on a ring road and densities of (a) $\rho=\unit[12]{veh.km}$, and
(b) $\rho=\unit[30]{veh.km}$. Plotted are the simulated data
(symbols), the theoretical distribution \protect\refkl{gs} (thick
solid line), and, for
the higher density, the Gaussian
approximations \protect\refkl{sigsGauss} (thin lines).
}
\end{figure}

\subsection{Gap distribution}

Figure \ref{fig:OVM-s} shows the simulated gap distributions for the
stochastic OVM for two densities and two values of the symmetry
parameter $\gamma$. For comparatively low densities
(Fig. \ref{fig:OVM-s} (a)), 
both the predicted and the observed
distributions are markedly  asymmetric. Furthermore, the direction of
interaction plays a role as well. For the limiting case of a car-following model
($\gamma=0$), the gap distribution is wider than in the symmetric
(momentum-conserving) case $\gamma=1$. Generally, there is a
good agreement between  the
theoretical expressions and the data. The only exception is the 
large-gap tail for symmetric forces at the lower density. In contrast,
for the car-following case $\gamma=0$, even the tails are reproduced
correctly (within statistical
fluctuations). The same aggreement has been found for the higher
density, irrespective of the value of $\gamma$.  This is remarkable since
additional assumptions have been necessary in Section \ref{sec:theo} 
to derive the  theoretical distributions
for the car-following case. Consequently, one would expect larger errors 
 compared to the isotropic case. 

\begin{figure}
\fig{0.45\textwidth}{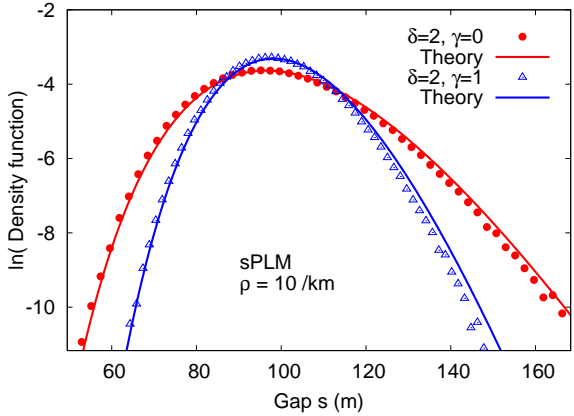}
\caption{\label{fig:power-s}Stationary gap distributions  for the
stochastic power-law model (sPLM) with the parameter and simulation settings 
specified in Table \protect\ref{tab:paramPow}.
}
\end{figure}

Now we investigate the influence of the densities on the form of the gap
distribution. Comparing Fig. \ref{fig:OVM-s}(a) with  
Fig. \ref{fig:OVM-s}(b), one may conclude that, when increasing the density, the
distributions become more and more symmetric.
Further simulations showed that the distribution becomes significantly
asymmetric if the single-particle kinetic energy $T=D\tau/4$ exceeds the 
effective potential energy $U(1/\rho)$ by at last one order of
magnitude. Specifically, for the situation of Fig. \ref{fig:OVM-s},
we have 
$T=\unit[1]{m^2/s^2}$ for all values of $\rho$ and $\gamma$
while the effective potential energy corresponding to  plot (a) is given by 
$U\sub{sOVM}= (1+\gamma) \, \unit[0.127]{m^2/s^2}$ and that of plot (b) by
$U\sub{sOVM}= (1+\gamma) \, \unit[95.0]{m^2/s^2}$. 

For sufficiently high densities when the standard deviation of the gap
distribution is much smaller than the average gap $1/\rho$, 
the Gaussian assumption
\refkl{sigsGauss} should become valid. To determine the
range of validity, we plotted 
the Gaussian approximation in the relevant Figure \ref{fig:OVM-s}(b),
in addition to the general distribution
\refkl{gs}.
For the case $\gamma=1$ corresponding to $\sigma_s^2=\unit[1.21]{m^2}$, 
the Gaussian approximation agrees nearly perfectly with the full
theoretical curve
(the curves overlap with no
visible difference).  For $\gamma=0$
($\sigma_s^2=\unit[2.42]{m^2}$), however, a significant difference
is found, but the distribution \refkl{gs} already displays a
significant skewness for this case. Further simulations showed that 
the Gaussian approximation is applicable
whenever the full distribution \refkl{gs} is sufficiently symmetric.

In order to evaluate the robustness of the predictions with respect to
different model types,  we have simulated the gap distributions for the
stochastic power-law model as well. The results are shown in 
Fig. \ref{fig:power-s}. Apart from minor deviations at the large-gap
tails, we found a remarkable agreement between theory and simulation.
Moreover, the predicted distributions
for the car-following case $\gamma=0$ and the symmetric case
$\gamma=1$ are significantly different both with respect to
variance and shape, which justifies the  particular specification \refkl{Ueff}
of the effective potential.

\begin{figure}
\fig{0.45\textwidth}{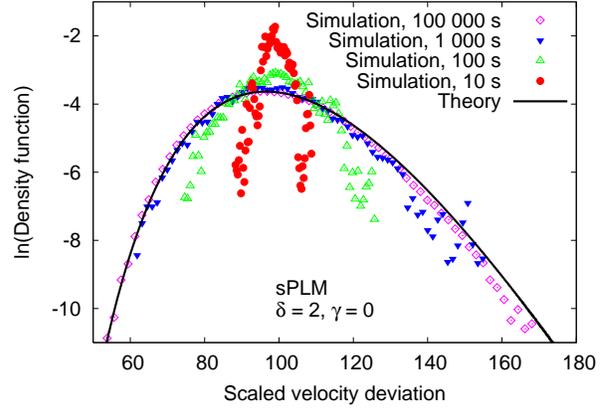}
\caption{\label{fig:relax-s}Relaxation of the  (initially
$\delta$-shaped) gap
distribution to its stationary distribution for the stochastic power-law model
with the parameters specified in Table \protect\ref{tab:paramPow}.
The figure compares simulation results at different points in time (symbols)
with the theoretical stationary distribution (solid line).
}
\end{figure}

Finally, we looked closer at the relaxation dynamics of the
initially $\delta$-correlated distributions, Eq. \refkl{IC}, towards the
stationary distributions. A very slow relaxation could be a possible reason
for the
deviations found sometimes at the large-distance tails of the gap
distributions.
In Fig. \ref{fig:relax-s},  we display 
snapshots of the evolution of the distribution for different simulation times. 
The results show that the relaxation time is considerable, particularly for low
densities. Moreover, the relaxation process is particularly slow at
the tails, so it may be a  plausible reason for the remaining deviations.

\begin{figure}
\fig{0.45\textwidth}{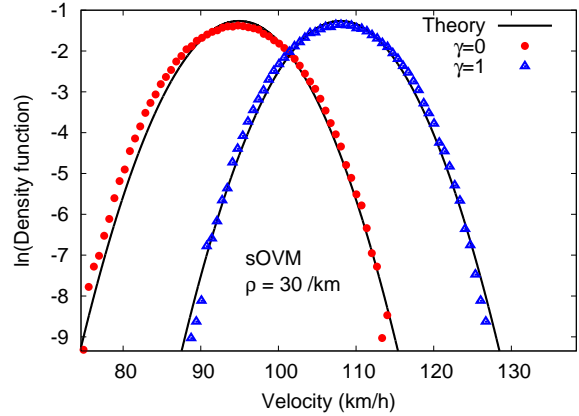}
\caption{\label{fig:OVM-v}Stationary velocity distribution  for the
stochastic OVM with the parameters specified in Table \protect\ref{tab:paramOVM}.
}
\end{figure}

\subsection{Velocity distribution}

In contrast to the gap distributions, the predicted velocity
distributions are always Gaussian. Moreover, the velocity variance
should
satisfy the fluctuation-dissipation theorem \refkl{FDT}. 
As a consequence, the variance may 
 neither depend on the density nor on the direction of the
interacting forces. Figure \ref{fig:OVM-v} shows simulated velocity
distributions for the stochastic OVM at the higher density corresponding to
Fig. \ref{fig:OVM-s}(b). One observes that, with the exception of
small but systematic deviations from the Gaussian shape for the
car-following case $\gamma=1$,
all theoretical predictions are fulfilled. For the lower density
$\rho=\unit[12]{/km}$ (not shown), the agreement was nearly perfect
for all values of $\gamma$.

In contrast to the gap distributions, the agreement  of the
velocity distributions 
improves when going from the car-following to the conservative case and when decreasing the density.
This can possibly be explained by the distance from the instability
point, see Eq. (\ref{linstab}). For the densities $\rho=\unit[12]{/km}$ and \unit[30]{/km},
the dimensionless distances 
$\ctrl=\tau_{\rm c}/\tau$ 
from the instability point are given by 
$\ctrl=0.001$ and $0.132$,
respectively, while we have $\ctrl=0$ for the conservative case.
Obviously, the agreement increases with the degree to which the
requirement $\ctrl \ll 1$ is satisfied.

\begin{figure}
\fig{0.45\textwidth}{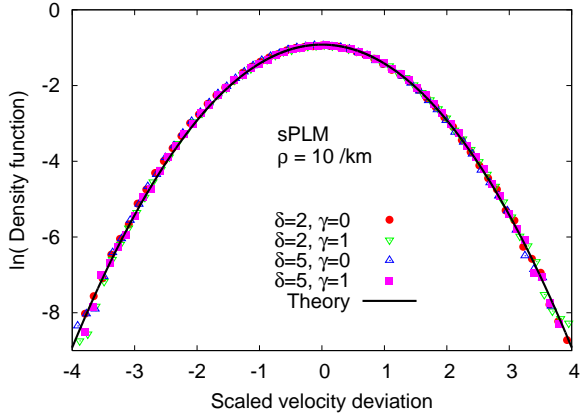}
\caption{\label{fig:power-v}Stationary velocity distributions  for the
stochastic power-law model with the parameters specified in Table \protect\ref{tab:paramPow}.
The velocity distributions have been normalized with respect  to the theoretical
expectation value $V$, see Eq. \protect\refkl{V-pow}, and the variance
\protect\refkl{FDT}.
}
\end{figure}

\begin{figure}
\fig{0.45\textwidth}{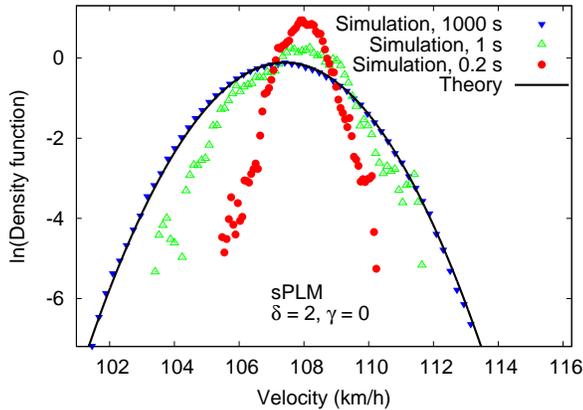}
\caption{\label{fig:relax-v}Relaxation of the velocity 
distribution to the Gaussian stationary distribution for the stochastic power-law
model with the parameters specified in Table \protect\ref{tab:paramPow}. Initially, all
particles had the same gaps $s_i=1/\rho$, and all velocities were
equal to the expectation value.
 The figure compares simulation results at different points in time (symbols)
with the theoretical stationary distribution (solid line).
}
\end{figure}

In Figure \ref{fig:power-v}, we have plotted the simulated
distributions for the stochastic power-law model for different values of the
density and the directional parameter $\gamma$. In each simulation, we have
normalized the distribution to the theoretical expectation value
\refkl{V-pow} and variance \refkl{FDT}, so we expect that all curves 
collapse onto each other in the ideal case. 
This collapse is, in fact, observed, thanks to values of 
$\ctrl$ below $0.12$ 
for all cases.

Finally, we investigate the relaxation process from the
$\delta$-correlated intial velocity distribution to the stationary
distribution. 
Figure \ref{fig:relax-v} shows that there is a significant scale
separation for the relaxation times: 
While the typical velocity relaxation time scale  is of the
order of seconds, it is of the order of hundred seconds for the
gap distribution.  After \unit[1\,000]{s}, even the tails of the
velocity distribution are perfectly equilibrated,
while for the gaps,  this takes longer by a factor of more than one hundred. 

We conclude that,  in contrast to the case of gap distributions, 
long relaxation times cannot explain possible differences 
between the theoretical and simulated velocity distributions,
as noticeable in Fig. \ref{fig:OVM-v}. These will be explained in the following.

\subsection{Kinetic energy and correlations}

One of the crucial assumption in the derivation of the gap and velocity
distributions of Sec. \ref{sec:theo} is the  assumption of zero
correlations, which requires that the system is far from any
instability, i.e. $\ctrl\ll 1$.
In classical  thermodynamic systems, it is well known and
theoretically understood \cite{landau-hyd}
that the energy contained in the fluctuations increases near a phase
transition resulting in ``critical opalescence'' and other observable 
phenomena. The same has been found in 
driven thermodynamic systems such as Rayleigh-B{\'e}nard convection or
electrohydrodynamic convection 
\cite{P-RMP-CrossHohenb-7v1-93} below the deterministic threshold,
which will be further discussed in Sec. \ref{sec:conclusion}. 

It is therfore very interesting to investigate the stochastic
properties of our driven particle-systems as a function of the
distance from threshold, i.e., varying the relaxation time from
$\tau=0$ to $\tau=\tau_{\rm c}$ or, equivalently, the control parameter
from $\ctrl=0$ to $\ctrl=1$.

\begin{figure}
\fig{0.45\textwidth}{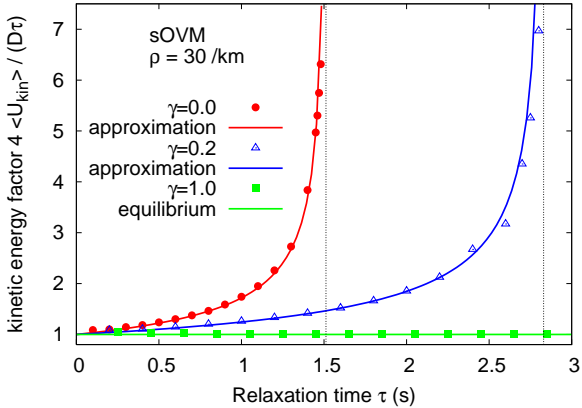}
\caption{\label{fig:tau}
Average kinetic energy per particle for different values of
the relaxation time and three values of the anisotropy parameter
$\gamma$. The amplification of the velocity variance with growing values
of $\tau$ reflects critical fluctuations close to the instability
point, see Eq. (\protect\ref{linstab}).
The energy has been plotted relative to the
theoretical value corresponding to the fluctuation-dissipation theorem 
(symbols). The solid curves display the function
$1/\sqrt{1-\ctrl}$ of the reduced control parameter \protect\refkl{ctrl},
and the thin vertical lines give the asympotics for $\tau\to\tau_c$
($r\to 1$).
}
\end{figure}

Figure \ref{fig:tau} shows the single-particle kinetic energy of the
fluctuations as a function of the relaxation time for several values
of the directional parameter $\gamma$. The kinetic energy has
been normalized to the value \refkl{kin} resulting from the
fluctuation-dissipation theorem \refkl{FDT}.
As in the physical systems mentioned above, we found significantly
increased, so-called  ``critical fluctuations'' near the linear threshold, which is
located at $\tau_{\rm c}=\unit[1.51]{s}$ for $\gamma=0$ and at $\tau_{\rm c}=\unit[2.83]{s}$
for $\gamma=0.2$ while no such threshold exists for $\gamma=1$. 

Finally, we compared the observed increase factor of the 
fluctation energy with the
function $1/\sqrt{1-\ctrl}$ of the scaled distance to the threshold
(scaled control parameter). The agreement was astonishing
for all investigated values of $\ctrl$ and $\gamma$.

\section{\label{sec:conclusion}Conclusion}
In this contribution, we have investigated  the statistical properties
of onedimensional dissipative driven
many-par\-ti\-cle systems violating the law ``actio = reactio''. Such
systems can represent, for example, vehicular traffic or queuing
systems with interactions. 

In the theoretical derivation, we have shown that such systems
approximately show a Hamiltonian statistic when inter-particle
correlations play no significant role. The theoretical predictions
were confirmed by simulations: Without a single free parameter to fit,
we quantitatively obtained the typical characteristic properties of Hamiltonian
systems such as velocity and gap statistics corresponding to a
canonical ensemble $\propto \mbox{e}^{-{\cal H}/\theta}$ when the
Hamiltonian ${\cal H}$ contains the usual contributions of kinetic and
potential energy as in physical Hamiltonian systems. Furthermore, the
velocity variance satisfied the fluctuation-dissipation theorem.
As only prerequisite, we have found that
the system must be far away from any instability
point. This is consistent with the theoretical requirement of
vanishing correlations, since collective instabilities
such as stop-and-go traffic correspond to highly correlated particles.

In the first moment, these results 
appear to be quite surprising.  For example, in traffic systems neither energy nor momentum
are conserved  during vehicle interactions, and the driving force keeps the system permanently far
from equilibrium.  While conservative systems conserve momentum and energy
in each single interaction, in the driven dissipative systems studied by us, the
additional relaxation term $(v_0-v_i)/\tau$ causes the
average velocity $V$ to relax to the ``free'' or ``desired'' speed $v_0$. 

However, even systems violating  the law ``actio=reactio''
may be mapped to effectively conservative systems by a Galilei
transformation. Defining motions relative to the stationary velocity,
\be
u_i=v_i-V
\ee
Eq. \refkl{SDEv} becomes
\be
\label{SDEu}
\abl{u_i}{t} = -\frac{u_i}{\tau} + f(s_i)-f(1/\rho)
+\gamma \left[ f(s_{i-1})-f(1/\rho) \right].
\ee
One sees that the constant terms $-f(1/\rho)$ and 
$\gamma f(1/\rho)$ resulting from the driving force and the relaxation
dynamics supplement the interaction forces by counteracting
forces -- irrespective of the value of $\gamma$ -- such as in
momentum-conserving systems.

One big difference of system not conserving momentum, however, remains when compared to
conservative system: 
The conservative many-particle system always behaves dynamically stable, while
the dissipative system potentially 
produces stop-and-go waves, when the linear stability condition~\ref{linstab} is not fulfilled.
According to computer simulations,
close to the instability point, the driven dissipative many-particle system
tends to produce correlations between distances and 
velocities and between successive
particles \cite{Schimansky-Geier-PRE98}.

This corresponds to pattern formation
phenomena that would not occur in conservative systems. 
Such kinds of pattern
formation phenomena have, for example, 
been investigated in fluid systems driven by thermal gradients
(Rayleigh-B{\'e}nard convection), coriolis forces (Taylor-Cou\-ette
flow), electrical fields (electroconvection), or concentration
gradients (binary-mixture convection), 
see Ref. \cite{P-RMP-CrossHohenb-7v1-93} for a review.
Moreover, the increase of thermal fluctations when approaching a
linear stability threshold from below has been investigated
theoretically and experimentally for the above systems
\cite{Rehberg-EHC-fluct,Ahlers-RBC-fluct,Treiber-Taylor-fluct,Rehberg-binMix2006}.
Near the threshold but in the regime of linear response, 
the fluctuations should increase according to a
power law, where the scaling exponents depend on the 
dimensionality and symmetry classes of the systems
\cite{Treiber-EHC-fluct}.
Specifically, if the fluid systems are 
quasi-onedimensional, 
their fluctuations typically increase proportional to $(1-\ctrl)^{-1/2}$ where
the reduced control parameter $\ctrl$ is defined
in analogy to  Eq. \refkl{ctrl}, with $\tau$ replaced by a suitable
driving force such as voltage or temperature gradient. Very near the
threshold, however, deviations have been observed empirically  \cite{Rehberg-PRL2000}.

It appears that in our case the fluctuations scale
proportionally to $1/\sqrt{1-\ctrl}$ as well. A
theoretical foundation and a closer investigation 
of the above Fokker-Planck equation near the instability point and
beyond will be subject of our future studies.





\end{document}